# Quantifying Multipolar Polarization

BSc Project BIBAPRO1PE

Christian Weidemann

chhhrz@gmail.com

IT University of Copenhagen

Supervised by Michele Coscia

## Abstract

Studying and understanding social networks is crucial for accurately defining ideological polarization, since they enable precise modeling of social structures. One of the limitations of many methods for quantifying polarization on networks is the assumption of a two-dimensional opinion space. This prevents accurate study of multipolar systems like multi-party political systems, where modeling more than two opinion poles is beneficial. Here, I experimentally compare methods for quantifying multipolar polarization on a network and find that the average pairwise distance extension of generalized Euclidean distance conforms to several desired properties, showing its advantages over other methods. This allows the study of multipolar polarized systems based on an empirically and intuitively good metric.

# Introduction

People often talk about how our society is getting more polarized, but to have any factual information addressing this sentiment, we must first agree on what we mean by polarization.

In political science, affective polarization is used to describe the tendency for positive and warm feelings about the in-group and negative feelings and sentiments towards the out-group [1], whereas ideological polarization refers broadly to a divergence of opinions and the tendency to communicate less with differently-minded people [2].

For a long time, ideological polarization has been measured with polling data [3], [4], [5], and theorists have devised numerous types, or senses, of polarization based on opinion distributions. Social networks provide a different approach and the ability to capture more nuanced aspects of polarization, like detailed structures describing how people communicate and associate.

Contrary to some, who use network polarization in a dynamic sense, a process which can be simulated [6], [7], [8], I use polarization as a static property of social networks and their node attributes; the tendency for people of similar stances to be connected, and for people of different stances to not be connected. Determining whether polarization is on the rise then becomes a question of measuring whether this property increases as the network changes through time.

Generalized Euclidean distance [10] is a solution to the node-vector distance problem [11], the problem of finding the distance between two vectors, whose entries are attributes of nodes on a network, in a way that is sensitive to network structure. This was shown by Hohmann, et al. (2023) [12] to be useful for measuring polarization, and to be sensitive to their proposed three components of polarization:

1. Opinion – whether people's stances tend to be divided rather than similar.
2. Dialogue – whether people communicate with others with similar stances to themselves, rather than opposing.
3. Opinion-dialogue interplay – whether communities associate with similar-minded communities, or if they prefer to engage others with opposing stances.

Hohmann, et al. (2023) [12] modeled stances as a bi-polar distribution on a spectrum from -1 to 1. Then they computed generalized Euclidean distance between two derived opinion vectors; one containing the positive stances and zeros elsewhere, and another with the absolute value of negative stances, and zeros elsewhere [12, p. 9].

A significant limitation with the metric is that it functions only for networks where opinions have this bi-polar distribution. This works for the United States where two parties dominate, but for multi-party political systems like Denmark, political leaning is more accurately quantified level of support for ~12 individual parties.

To address this limitation, I propose an extension to generalized Euclidean distance for measuring multipolar polarization; the case of more than two opinion vectors. I use synthetically generated network data and compare several candidate methods through experiments that are not covered by the three components of polarization.

For a comparison method not based on generalized Euclidean distance, I use a substantially different metric, Total Variation, proposed by Martin-Gutierrez, et al. (2023) [16].

# Definitions

A **Network**, also called a graph, is a set of nodes that are interconnected by a set of edges – pairs of nodes. Visually, nodes are represented as dots, with edges being lines connecting two dots. Networks can thus be summarized by the number of nodes $|V|$ and the number of edges $|E|$. For simplicity, I will treat networks as (1) undirected; the edge $(a, b)$ is equivalent to $(b, a)$, (2) unweighted; edges have no weight (or other features), and (3) as having no duplicate edges. A **path** is a set of edges (and their nodes) connecting one node in the network to another. In social networks, nodes often represent people, with edges being relationships between them. Therefore, I use the terms 'node', 'individual', and 'person' interchangeably.

**Network opinion,** or **opinion** for short, is a vector $o$ of length $|V|$ which holds a number, called **stance**, in the interval $[0,\ 1]$ for each person, describing the degree to which they subscribe to the opinion. Each node then has $|O|$ stances, the number of network opinions.

For example, when a network's nodes are voters, an opinion could be made up of each voter's stance towards candidate A.

**Orthogonal opinions** are a pairwise orthogonal set of opinions, defined $\sum_i \sum_j o_i \cdot o_j = 0 \text{ and } o_i \neq \vec{0}$, i.e., the dot-product between any two orthogonal opinions is zero and orthogonal opinions are not the zero vector. An opinion can be **unique** concerning a set of nodes, meaning the nodes have stance 1 for that opinion and 0 for all others.

A **community** is loosely defined as a set of people who hold similar stances and/or have a higher probability of being connected, relative to network nodes outside the set.

**Generalized Euclidean distance** [10] between opinions $a$ and $b$ on a network is defined:

$$\delta_{G,O}(a,b) = \sqrt{(a-b)^T L^\dagger (a-b)}$$

Where $L^\dagger$ is the Moore-Penrose pseudoinverse of the network Laplacian.

For a network, the Laplacian matrix $L$ is equal to $D - A$, where $D$ is the degree matrix and $A$ is the adjacency matrix of the network.

The Laplacian has many useful properties, including describing the discrete flow of heat between nodes, given a temperature at each node. The inverse of the Laplacian would be related to the time taken for heat to diffuse to equilibrium, but since the Laplacian is not invertible, the Moore-Penrose pseudoinverse is used [10].

Due to the properties of this matrix, generalized Euclidean distance is proportional to the square root of the time it takes for heat to flow from temperatures $a$ to temperatures $b$ through the network [12, pp. 24–25]. For the same reasons, generalized Euclidean distance polarization is lower when the two opinions are more similar and when paths connecting nodes who are unique to each opinion are more numerous, shorter, and go through nodes with similar stances.

The metric is not normalized, since it lacks a well-defined maximum [12, p. 7].

# Candidate Methods

I use 5 methods for measuring multipolar polarization of a network: (1) Average Pairwise Distance, (2) Average Distance to Mean, (3) Principal Component, (4) Multidimensional Scaling, and (5) Total Variation.

The first four methods are candidates for multipolar generalized Euclidean distance. Average Pairwise Distance and Average Distance to Mean are conceptually and computationally simple methods that rely on averaging several generalized Euclidean distances.

Principal Component and Multidimensional Scaling, on the other hand, are dimensionality reduction techniques, which attempt to meaningfully reduce the dimensionality of the opinion space to one dimension, one vector. Principal Component linearly does this transformation, whereas Multidimensional Scaling does this in a network-dependent nonlinear fashion. As in generalized Euclidean distance, the proposed methods use the Moore-Penrose pseudoinverse of the network Laplacian to solve the node vector distance problem [11] for some specified vectors.

The fifth method, Total Variation, is not based on generalized Euclidean distance, but is included as a state-of-the-art comparison metric for multipolar polarization.

## Average Pairwise Distance (APD)

This approach, proposed by Hohmann, et al. (2023) [12, pp. 22–23], simply averages the generalized Euclidean distances between each unique pair of opinions:

$$APD(G,\ O) = \frac{1}{\binom{|O|}{2}} \sum_{i}^{|O|} \sum_{j>i}^{|O|} \sqrt{(o_i - o_j)^T L^{\dagger} (o_i - o_j)}$$

Where $G$ is the network and $O$ is the matrix of network opinions on columns.

## Average Distance to Mean (ADM)

This method consists of averaging the generalized Euclidean distance between each opinion and the arithmetic mean over opinions:

$$ADM(G,\ O) = \frac{1}{|O|} \sum_{i=1}^{|O|} \sqrt{(o_i - \overline{o})^T L^{\dagger} (o_i - \overline{o})}$$

Where $\overline{o}$ is the arithmetic mean of network opinions. A possible variation of this could be to use the squared deviation from the mean, but I don't use this due to a lack of justification for heavily emphasizing extreme opinions.

## Principal Component (PC)

Principal component analysis (PCA) is a popular dimensionality reduction technique. It is often used to reduce high-dimensional features to a few significant ones, or to visualize high-dimensional datasets using the first two principal components. However, here I use it to reduce opinions down to a single dimension, the first principal component. Then stances are projected onto it to give a single vector $o_{PC}$, and polarization is computed:

$$PC(G,\ O) = \sqrt{o_{PC}^T L^{\dagger} o_{PC}}$$

This formulation is equivalent to the generalized Euclidean distance between the zero vector and $o_{PC}$.

Since principal components are unit vectors (their length is one), in the case where all stances are also unit vectors, projecting stances onto the first principal component guarantees that the transformed stance is in [0, 1], meaning $o_{PC}$ would satisfy the definition of an opinion. As a research design decision, normalization of stances (transforming them to length one) might be sensible depending on the data domain. It would be sensible to do, if a person's stances represent probabilities of them belonging to each opinion, or if they should be interpretable as such. In that case, each person has the same total contribution to network opinions. On the other hand, the normalization of stances is inappropriate when a person's stance toward one opinion shouldn't limit their other stances, e.g., when stances represent independent preferences, choices, or beliefs. Since this project has no specific data domain, I use the most general option, not normalizing stances, although they happen to be unit vectors in some experiments.

## Multidimensional Scaling (MDS)

Multidimensional scaling (MDS) is a family of nonlinear dimensionality reduction techniques that seek to preserve (relative) inter-data point distances as well as possible, in the lower-dimensional space [13]. It is similar to PCA in that it computes weights for transforming the data, except instead of scaling features linearly, the position of each data point in the lower-dimensional space is optimized [13]. I use the Sci-kit Learn implementation of metric absolute MDS, which uses the squared error loss function "stress" [13], [14]:

$$Stress(o_{MDS}) = \sum_{i<j} d_{ij}(o_{MDS}) - \hat{d}_{ij}(o_{MDS})$$

Where $o_{MDS}$ is the transformed opinion vector, $\hat{d}_{ij}$ is the MDS transformation on distances between stances $i$ and $j$ and $d_{ij}(O)$ is the distance according to a user-specified distance function. Euclidean distance is commonly used for this, but custom functions can be used. I experimented with using effective resistance distance [15], which takes all possible paths between the two nodes into account [12, p. 26], but this is independent of stance information by default. Therefore, Euclidean distance between stances is used instead.

When using MDS in practice, one can try many random initializations of the weight vector and use the embedding whose stress was lowest. However, since I'm interested in the general behavior of the method, I instead report the mean and 95-percentile confidence interval of polarization across 100 random initializations of weights, uniform in $[0,\ 1)$.

## Total Variation (TV)

To compare GED-based methods to a different state-of-the-art approach, I use total variation (TV), proposed by Martin-Gutierrez, et al. (2023) [16] for measuring multipolar polarization. TV is equal to the trace of the covariance matrix of the matrix O, with opinions as columns:

$$TV(G,\ O) = tr(Cov(O))$$

In other words, as follows from the definitions of covariance and trace of a matrix, TV equals the sum of opinion variances:

$$TV(G,\ O) = \sum_{i}^{|O|} Var(o_i)$$

Martin-Gutierrez, et al. (2023) showed that if opinions have a maximum magnitude of 1, the total variance of opinions also has the maximum value of 1 [16, p. 9]. Therefore, since variance can't be negative, TV is constrained to [0, 1], unlike the previously presented methods.

This method only depends on opinions, not the network structure. It assumes that all polarization-relevant network information is implicit in the opinion vectors. This was a valid assumption for Martin-Gutierrez, et al. (2023), since they used a network structure-dependent opinion inference technique. Since I don't do so, TV will only depend on my synthetically generated opinions. Even without inferring opinions from network structure, using TV for realistic social networks is sound, since in those networks, structure is presumably associated with opinions and vice-versa. This assumption doesn't hold for all synthetic experiments, since some change network structure independently of opinions.

## Network Generation

For experimentation, I use four simple network types. Chains, also called path graphs, are networks where each successive node is connected to the previously added node, at the end of the chain, as shown in Figure 1. Complete networks, or cliques, have each node directly connected to every other node. The third local-scale network I use has a simple community structure; complete networks are connected by two edges between neighboring pairs of nodes from each network.

The minimum $|V|$ in these networks is dictated by the third type, since for $n$ communities, fewer than $3n$ nodes would compromise the community structure; when $|V| < 3n$, i.e., when there are fewer than 3 nodes per community, some nodes are no longer more connected to their community than other communities. $|V|$ is incremented by $n$ to keep communities balanced.

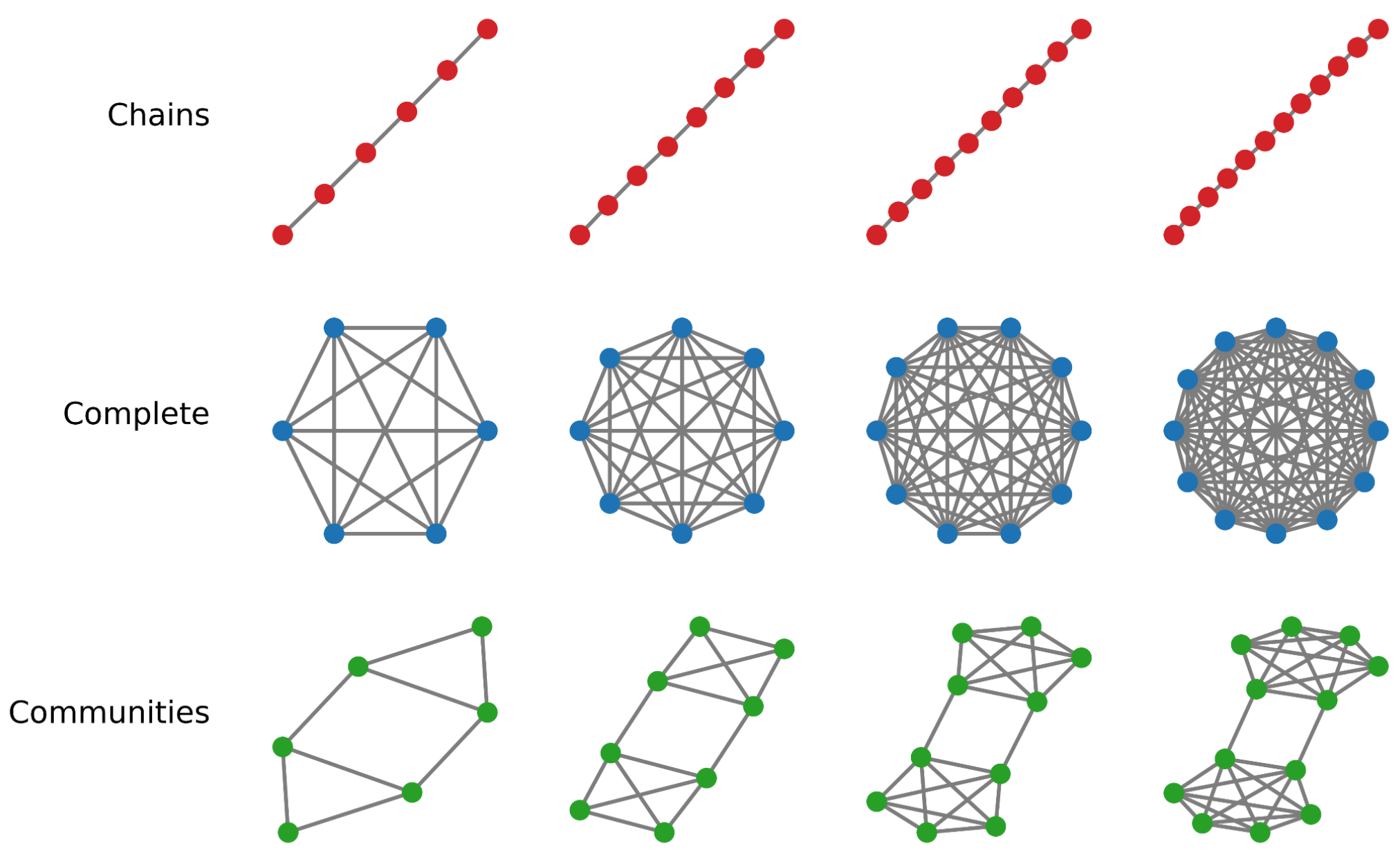

Figure 1: Three local-scale network types, separated by color: chains (red), complete network (blue), and communities (green). For community networks $n = 2$. The X-axis is the number of nodes $|V|$, going from 6 to 12 in increments of 2.

The stochastic block model (SBM) is a random generative model for networks [17], defined by $|V|$, a partition of nodes into $n$ communities and $P$, a $n \times n$ matrix with $P_{ij}$ being the probability of an edge between any two nodes in communities $i$ and $j$ [17]. I choose $|V|$ s.t. each community has 100 nodes. I use a simple version of this model, called planted partition [18], [19], where $P$ is only defined by the probabilities of each edge within a community being generated ($p_{in}$) and each edge between communities being generated ($p_{out}$):

$$P_{ij} = \begin{cases} p_{in} & \text{if } i = j \\ p_{out} & \text{otherwise} \end{cases}$$

This results in $P$ being symmetric, with all diagonal entries being $p_{in}$ and all off-diagonal entries being $p_{out}$. I use $p_{in} = 0.1$ and $p_{out} = 0.01$. Choice of edge probabilities is somewhat arbitrary, but follows some requirements to mimic real-world networks with distinguishable communities:

- $p_{in} > p_{out}$ since the social networks of interest to me are assortative, that is to say, people are more likely to be connected if they have similar stances.
- $p_{in} > (1-n)p_{out}$ ensures there are more edges within a community than going out of it.

Because the SBM is probabilistic, successively generated networks will be different. Therefore, for accurate reporting and comparison, polarization of the network is averaged across 10 random initializations, unless otherwise stated, and 95-percentile confidence intervals are shown. An exception to this is MDS polarization of SBM networks, where each of the 100 random MDS initializations also always uses a uniquely generated SBM.

# Synthetic Experiments

I subject each method to a series of experiments with synthetically generated networks to verify their conformity to expected and desired properties. For each experiment I present a scenario that changes the network and (1) describe how polarization should behave and why, (2) test how the methods behave with the scenario implemented on a local scale with small chain, complete and community networks, (3) test how the methods behave with the scenario implemented on large-scale SBM networks, and based on these tests, especially community and SBM networks, (4) conclude whether each method fulfills desired properties.

The purpose of the local-scale tests is to illustrate the methods' behaviors in relatively simple cases, giving some intuition for how they function and why. Say, large-scale real-life networks can be seen as a combination of hundreds of local structures, interacting to form a larger whole. Then, behavior on a large scale may be interpreted as the aggregation of behaviors on individual local structures. Using this bottom-up approach, the experiments are designed to yield understanding through interpretation of results, as well as empirical comparisons for decision making.

In interpreting results, I describe methods as "decreasing", "increasing", or "staying constant" as shorthand for polarization change according to the method, as the corresponding network changes.

## Adding nodes with unique opinions

This scenario consists of adding people to the network, whose stances differ from everyone else. The initial state of the network is with each person having a unique orthogonal opinion (where their stance is 1 and everyone else's is 0). Then, new people are added, also with unique orthogonal opinions.

### 1. Desired behavior

The number of opinions, all other things equal, should not affect polarization. Although more opinions mean more information, the state of polarization doesn't change – it is maximal, relative to the network and community structure. Polarization may decrease slightly due to the network becoming more diverse and multifaceted. Therefore, polarization should generally stay the same or slightly decrease.

## 2. Local scale

Adding nodes in the local-scale test is illustrated in Figure 2. Each node has a unique color, corresponding to its unique opinion. The community networks are arbitrarily decided to have 2 communities, and the $|V|$ is increased from 6 to 12 in increments of 2.

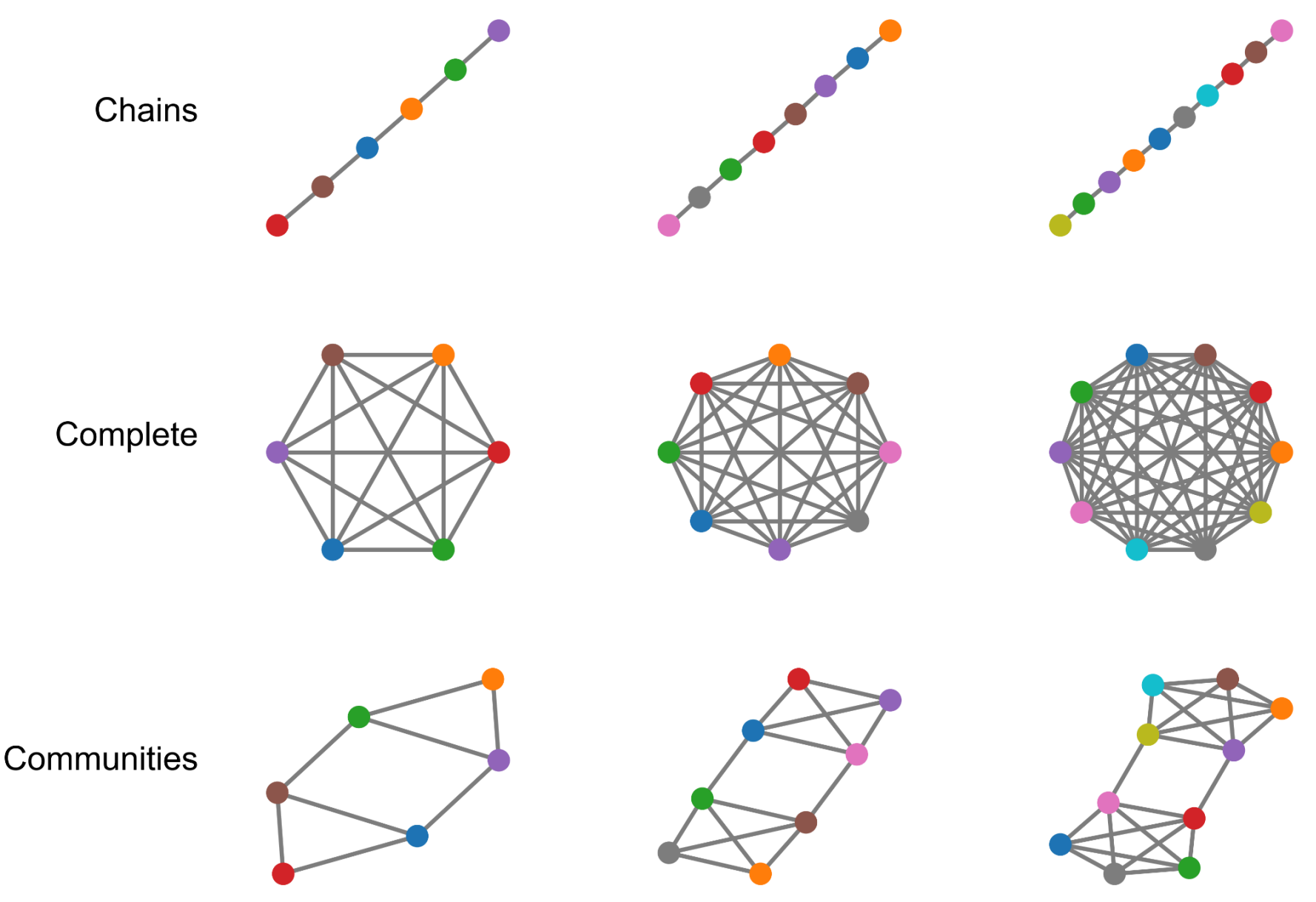

Figure 2: Examples of local-scale networks with each node having a unique orthogonal opinion, represented by node color. Network types are on rows, with community networks having 2 communities. Number of nodes $|V|$ are on columns, with 6, 8, and 10 in each network.

Figure 3 shows polarization values for each method across the three local-scale experiments. Since nodes have unique orthogonal opinions, APD and ADM change with average path length between any two nodes, increasing in chains but decreasing in complete and community networks. PC increases in chains and stays constant in communities, also over many more than 12 nodes, though doing so irregularly. The orientation of the first principal component is simply too sensitive to changes in the sparse high-dimensional space when opinions lack linear correlations. This irregularity is dampened in complete networks, where PC decreases smoothly, since path lengths don't change with new nodes. This same pattern happens for MDS, irregularly increasing in chains and communities, but staying constant in complete networks. TV is constant at its maximum value of 1, since opinions are maximally extreme and unique across nodes.

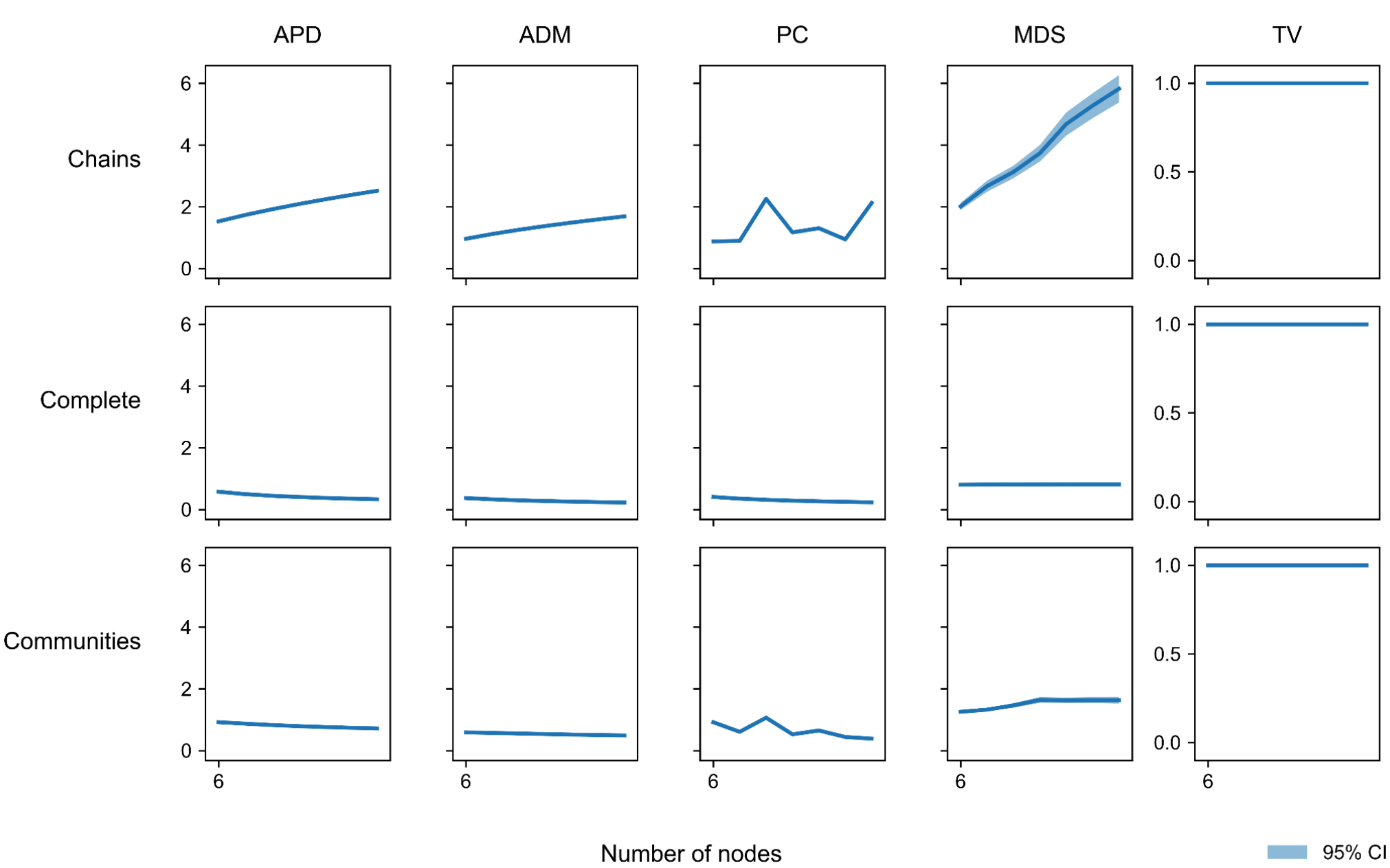

Figure 3: Polarization of local-scale networks when adding nodes each with a unique orthogonal opinion. Multipolar polarization methods are on columns, and network types are on rows. Y-axis shows polarization according to the corresponding method, while the x-axis shows the number of nodes $|V|$. The shaded areas are 95-percentile confidence intervals.

## 3. Large scale

To test the same principle on large-scale networks, I construct SBM networks with each new community of 100 nodes getting a unique orthogonal opinion. The number of communities $n$ goes from 2 to 6, as illustrated in Figure 4.

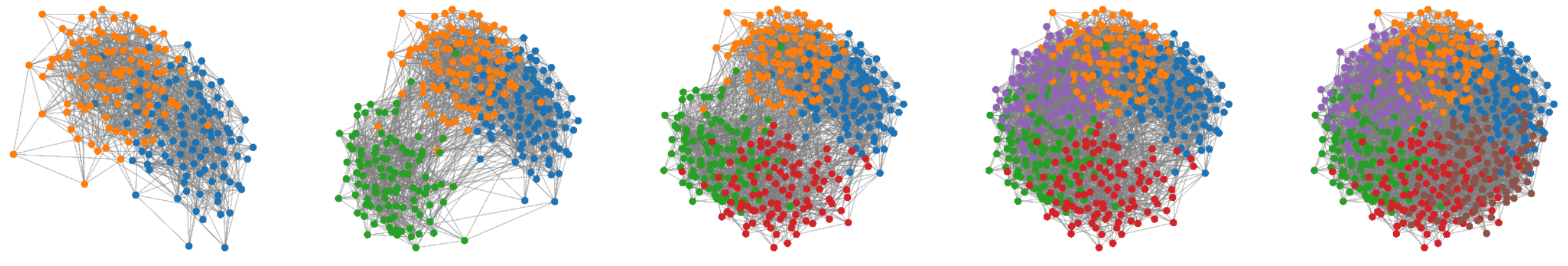

Figure 4: Example of generated SBM networks with 2 to 6 communities. For all networks $p_{in} = 0.1$, $p_{out} = 0.01$ and the number of nodes per community is 100. Node color represents stances, with each community having a unique orthogonal opinion.

The networks are created in the opposite direction, by first generating with $n = 6$, and iteratively removing a community and its corresponding out-edges until $n = 2$. This algorithm keeps edge probabilities constant as communities are added. Polarization is still expected to decrease, since the ratio of edges going between communities to edges within communities increases with $n$, reducing path lengths.

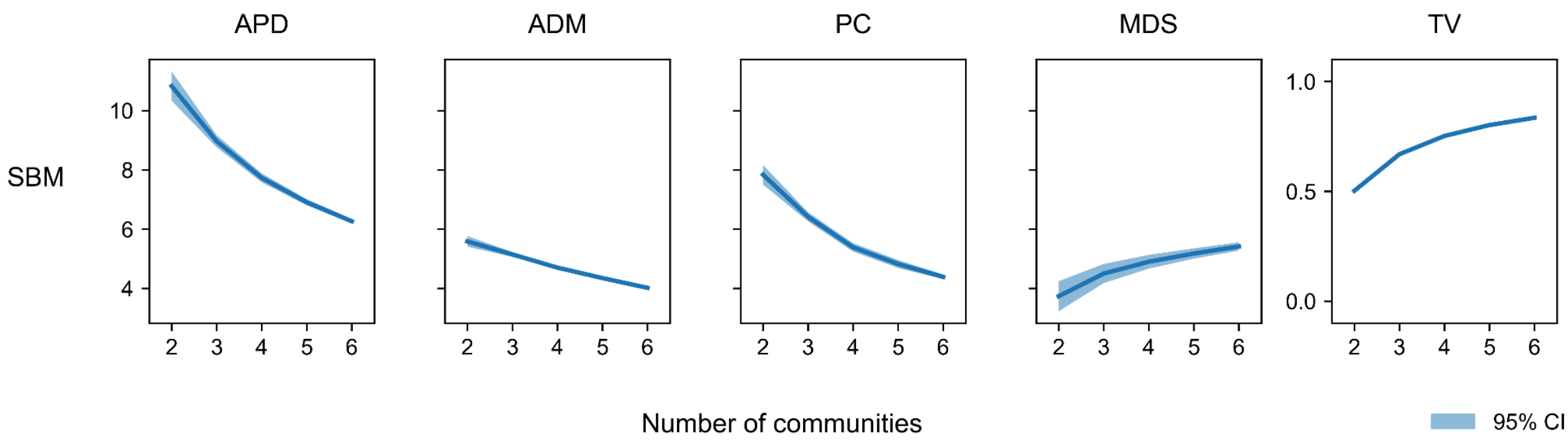

Figure 5: Polarization of SBM networks when adding communities with unique orthogonal opinions. Multipolar polarization methods are on columns. Y-axis shows polarization according to the corresponding method, x-axis shows number of communities $n$, each of 100 nodes. Shaded area is 95-percentile confidence interval.

Polarization decreases as expected for APD, ADM, and PC. MDS behaves differently, increasing with added communities. Since nodes in the same community have the same transformed stance in $o_{MDS}$, there are effectively two stances at $n = 2$, which MDS can space perfectly at a distance of $\sqrt{2}$. As $n$ goes up, new communities seek to keep the same stance distances to other communities, but this is not possible on the one-dimensional $o_{MDS}$. As a result, stress goes up and stances are spread further apart, increasing polarization. TV increases with new communities, since adding a new opinion dominates the small variance decrease of existing opinions.

### 4. Conclusion

APD and ADM correspond to desired behavior, decreasing or staying constant with added unique orthogonal opinions in the community and SBM networks. PC corresponds to desired behavior, staying constant in community networks and decreasing in SBM networks. MDS and TV don't behave as desired, increasing with $n$.

## Neutral nodes as a different community

This scenario models the research design decision to interpret neutral nodes – the nodes whose stances are between extremes – to instead have a unique opinion, orthogonal to existing ones. This may occur in 3-party systems with two extremes, where whether to differentiate members of the neutral party with their own opinion, rather than as half-agreeing with the two extreme parties' opinions, is a research design choice.

### 1. Desired behavior

I'm modeling specifically the case when the choice is arbitrary and up to interpretation, since the neutral nodes on one hand are structurally between the extremes, but on the other hand, distinctly different and separate from those communities. Because the choice is arbitrary, interpreting neutral nodes having a unique orthogonal opinion should have little or no effect on polarization.

### 2. Local scale

In the local-scale test, nodes are initially split evenly into 3 communities, two opposite extremes and a neutral third. The two communities of extreme nodes each have unique opinions, orthogonal to each other, whereas the neutrals have stances that are the average of the two extremes, $[0.5, 0.5]$. The neutral nodes are then changed to have unique orthogonal opinions as well, as shown in Figure 6.

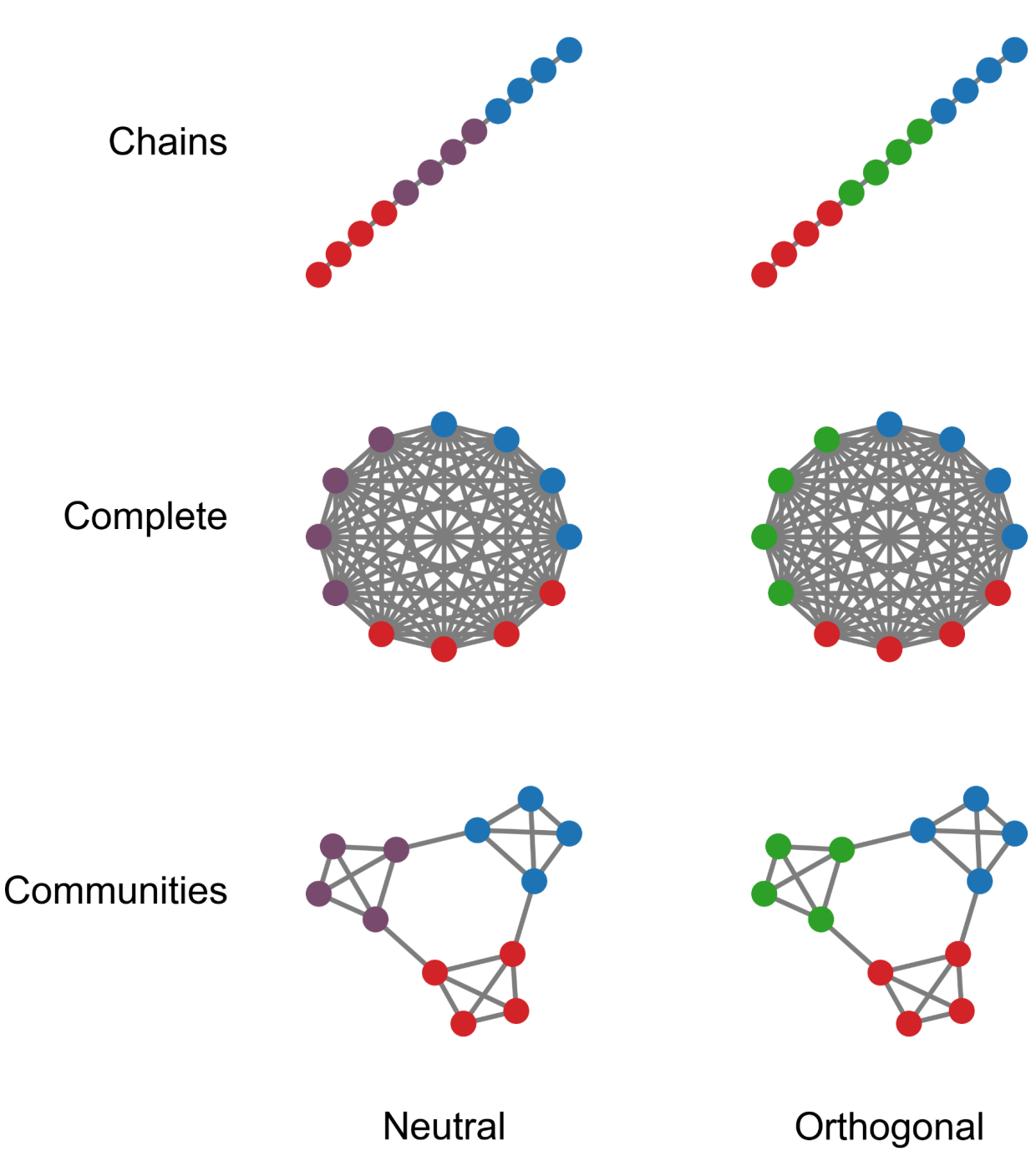

Figure 6: Examples of $|V| = 12$ local-scale networks with a neutral community (left) or a unique orthogonal community (right). Stance is represented by node color, with purple being the average of blue and red. Network types are on rows, with community networks having 3 structural communities.

I evaluate the effect on polarization across networks with $|V|$ increasing from 9 to 18 in increments of 3, as shown in Figure 7. In chains, APD polarization change is negative and decreases with $|V|$, even though pairwise opinion distances stay at $\sqrt{2}$, since in one case opinions can only flow one-way away from the ends of the chain, whereas in the other, an opinion can flow both ways from the center, reducing inter-opinion path lengths (lengths of paths connecting nodes who have different unique opinions). This distinction doesn't exist in complete and community networks, so APD remains unchanged here. ADM polarization change is negative in chains as well, but slightly positive in complete and community networks, since one community being neutral brings the distances between opinions and their mean down. PC polarization change in chains is negative and decreasing with $|V|$ while wildly fluctuating, while it is zero in complete and community networks. This happens because the orientation of the first principal component is arbitrary in networks with unique orthogonal opinions, but has a large effect only in the chain networks. MDS polarization change in all network types is positive since stance distances increase. It fluctuates in chains and communities, but not in complete networks, since only in those is the path from all nodes to other communities equally long. TV polarization change is positive, both since a new opinion is added and because opinion variances increase.

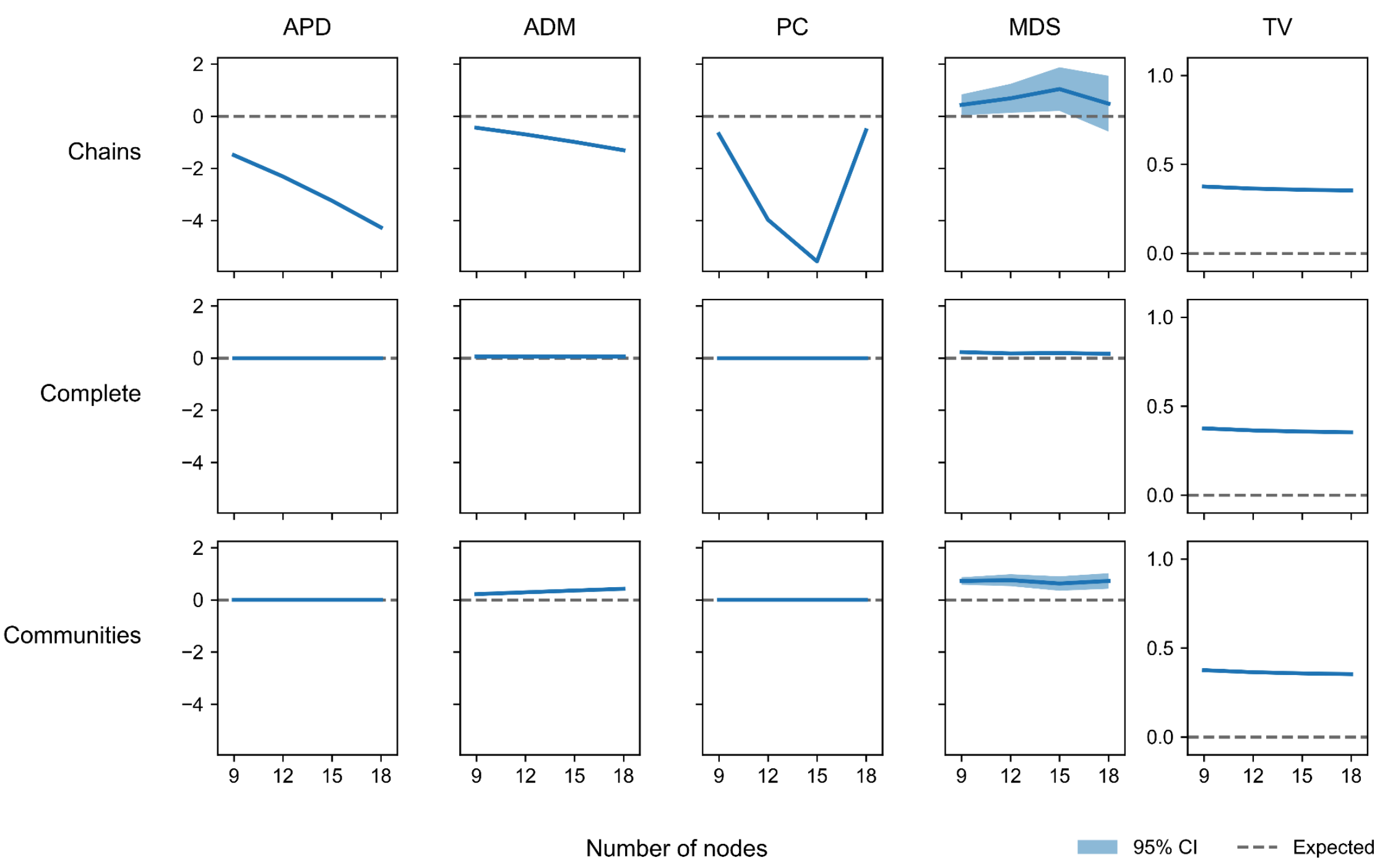

Figure 7: Polarization change going from neutral to orthogonal opinions on local-scale networks. Multipolar polarization methods are on columns, and network typesare on rows. Y-axis shows polarization change according to the corresponding method, x-axis shows the number of nodes $|V|$. The shaded areas are 95-percentile confidence intervals. The dashed gray line at y = 0 indicates how polarization is not expected to change.

## 3. Large scale

In the large-scale test, 3 communities – each of 100 nodes – are connected to each other. The opinions are changed like in the local-scale test, with the middle community initially being neutral, but then changed to have a unique opinion, orthogonal to the other two, as shown in Figure 8.

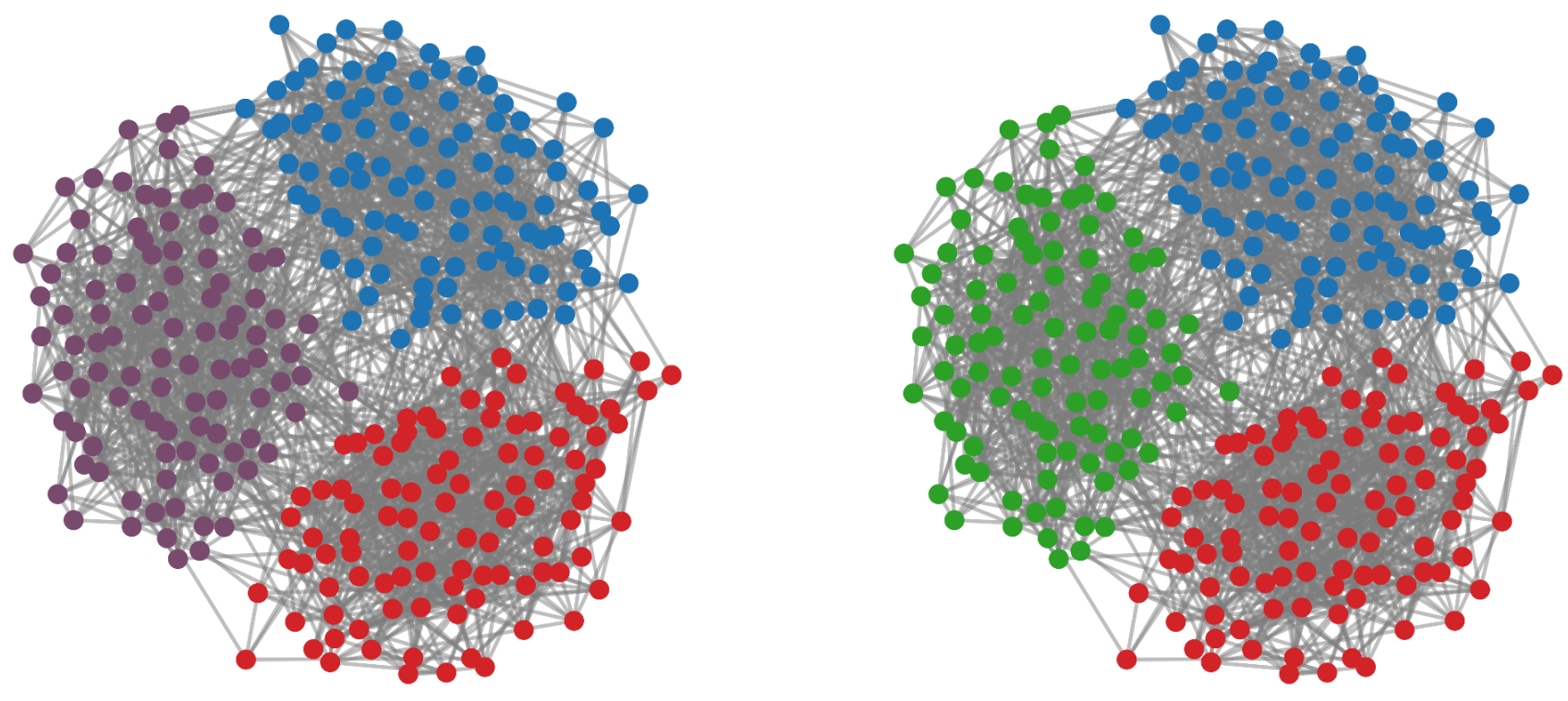

Figure 8: Example of generated SBM networks with a community going from neutral stances (left) to having a unique orthogonal opinion (right). Stance is represented by node color, with purple being the average of blue and red. The number of nodes per community is 100, $p_{in} = 0.1$ and $p_{out} = 0.01$.

The polarization change on SBM networks is shown for each method in Figure 9. APD and PC conform to desired behavior and stay constant, whereas ADM, MDS, and TV increase. ADM and MDS increase for the same reasons as in local complete and community networks. TV again increases as in the local tests, since an opinion is added, and opinion variances increase.

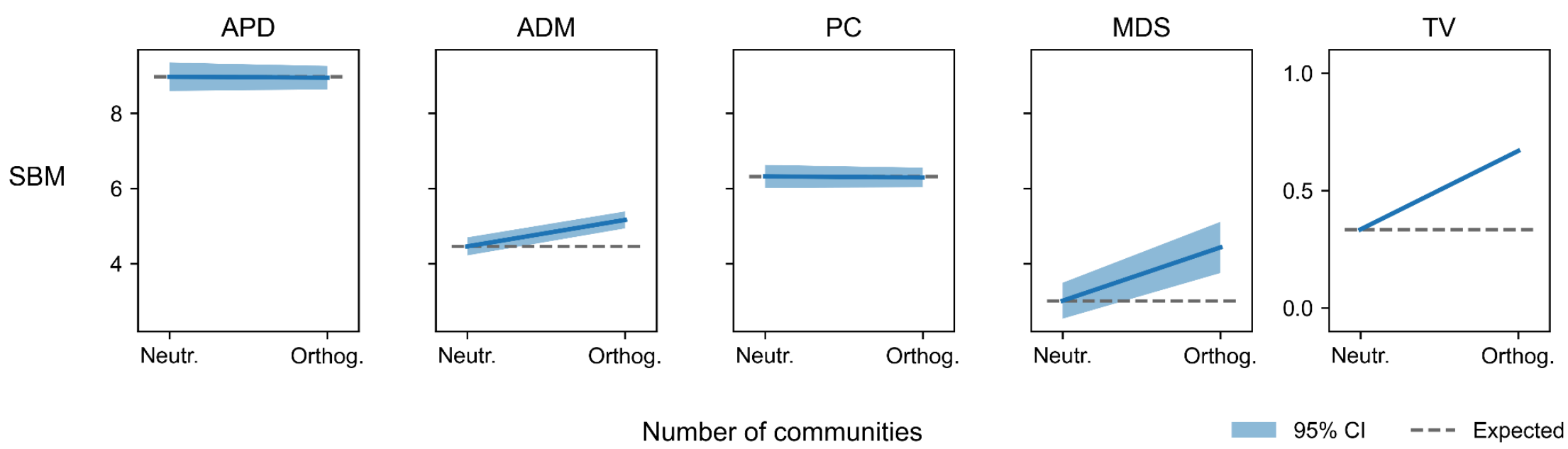

Figure 9: Polarization of networks with a neutral compared to an orthogonal opinion. Multipolar polarization methods are on columns. Y-axis shows polarization according to the corresponding method, x-axis represents the change from neutral to orthogonal opinion. Dashed gray line indicates the level of polarization with a neutral community, which is expected to stay the same for when it has a unique orthogonal opinion. Generated from 100 SBM randomizations, except for MDS which used 500 randomizations.

### 4. Conclusion

APD and PCA conform to desired behavior, staying constant when a neutral community is arbitrarily interpreted as having a unique orthogonal opinion, whereas ADM, MDS, and TV don't.

## Correlating two opinions

This scenario tests polarization behavior as the correlation between two of three opinions increases. Initially, three orthogonal opinions, $o_A$, $o_B$, and $o_C$, are each unique to a community. Then $o_A$ and $o_B$ gradually move towards each other in opinion space, until $o_A = o_B$. In the real world, this could play out in a political system of three parties, where two seek to gain the favor of each other's voters, resulting in mutual ideological attraction, until eventually being equally appealing to their combined voter base.

### 1. Desired behavior

Polarization should decrease, since stance diversity goes down as two-thirds of the population move from having minority to majority stances.

### 2. Local scale

Sizes of local-scale networks are fixed at $|V| = 12$, and instead the changed parameter is the angle $\theta$ between $o_A$ and $o_B$, ranging from 90° to 0° in increments of 10°, as illustrated for local-scale networks in Figure 10. The opinion vectors are rotated 5° toward each other at each step, keeping the magnitude constant. The experiment is called correlation even though angles are used, since $\cos(\theta) = corr(o_A, o_B)$ and $\cos(0) = 1$, so maximizing correlation is equivalent to minimizing the angle between opinions.

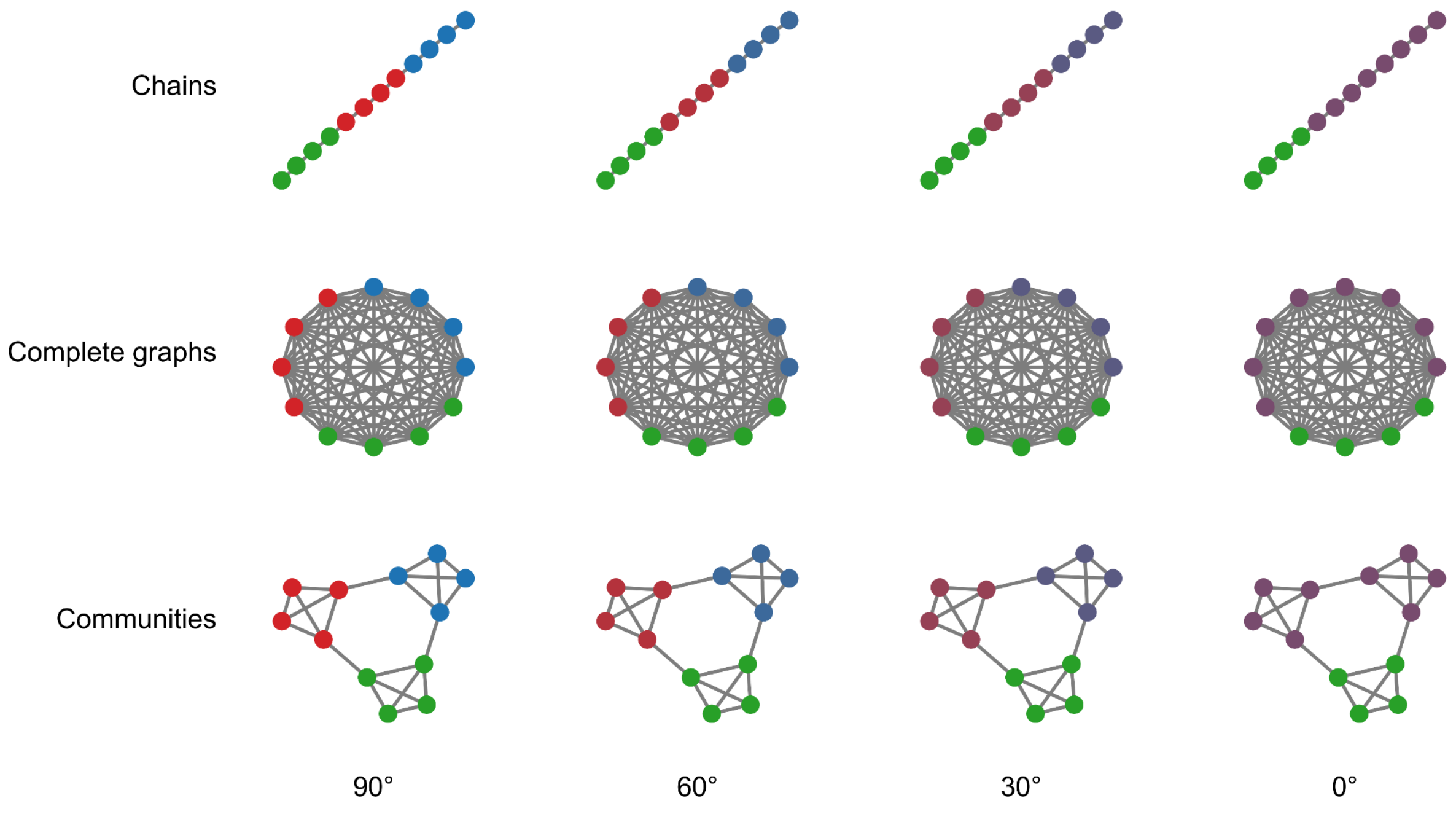

Figure 10: Examples of local-scale networks with 3 communities, each with an initially unique orthogonal opinion, represented by node color. The angle $\theta$ between $o_A$ and $o_B$, shown for each column in the bottom row, is reduced until $o_A = o_B$. Color indicates stance, with red and blue nodes turning purple as their stances get more similar.

Methods are evaluated on local-scale networks, as shown in Figure 11. APD decreases in all three networks as $\theta$ goes to 0, since the average pairwise Euclidean distance between opinions decreases while network structure stays constant. ADM also decreases in all network types, since the mean of opinions moves closer to $o_A = o_B$ as $\theta$ goes down. PC increases in all local-scale networks when $\theta < 90°$ and MDS stays generally constant in all three networks, but fluctuates since the three opinions can't be spaced evenly on the one-dimensional $o_{MDS}$. TV decreases because the variances of $o_A$ and $o_B$ decrease.

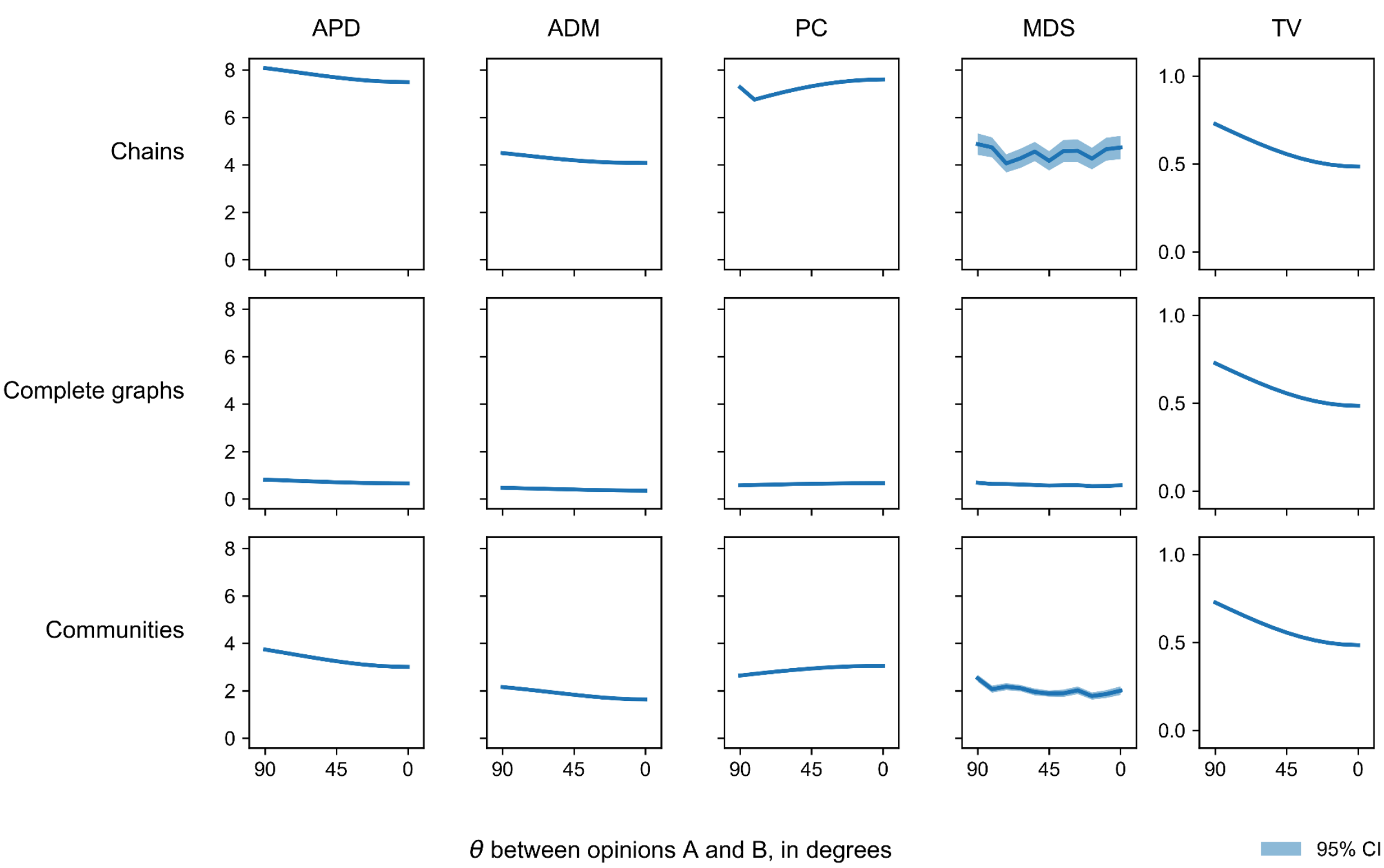

Figure 11: Polarization as the angle $\theta$ between $o_A$ and $o_B$ goes from 90° to 0° on local-scale networks. Multipolar polarization methods are on columns, and network types are on rows. Y-axis shows polarization according to the corresponding method, x-axis shows $\theta$. The shaded areas are 95-percentile confidence intervals.

### 3. Large scale

The large-scale test functions exactly as the local-scale tests, except with SBM networks with 3 communities of 100 nodes. $\theta$ ranges from 90° to 0° in increments of -10°, as illustrated (with larger increments) in Figure 12.

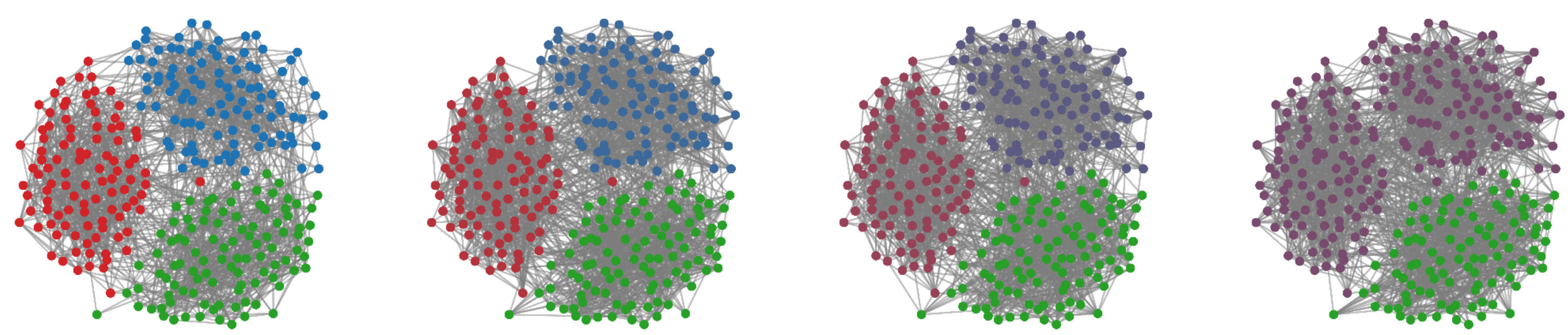

Figure 12: Example of generated SBM networks with the angle $\theta$ being 90° (left), 60°, 30° and 0° (right). Stance is represented by node color, with purple being in between blue and red.

For the same reasons as the local tests, in SBM networks APD, ADM, and TV decrease,

while PC increases and MDS stays constant and fluctuates, as shown in Figure 13.

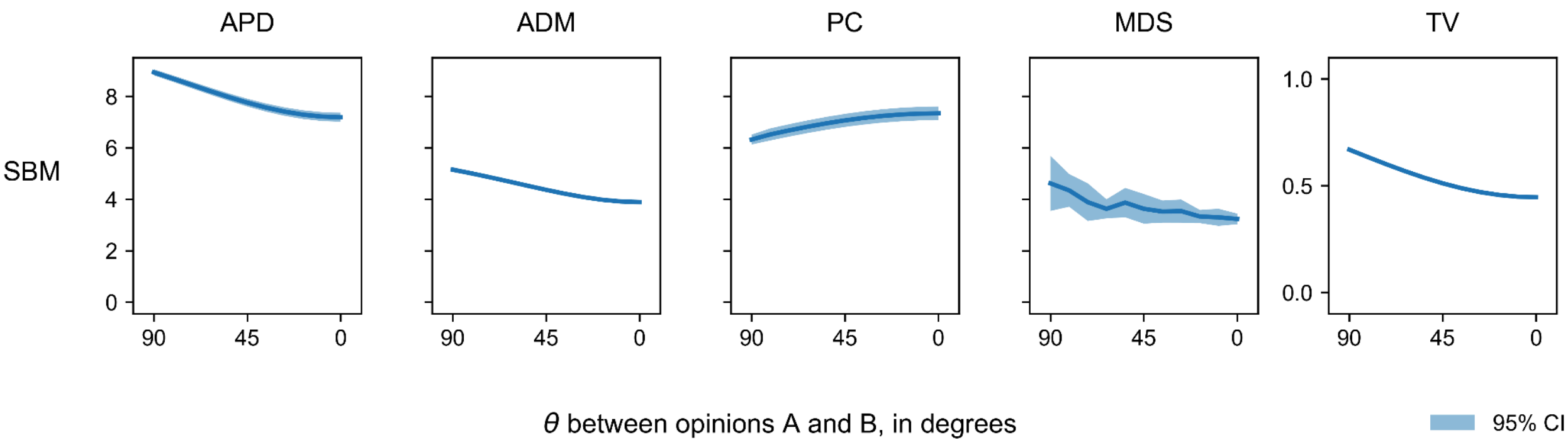

Figure 13: Polarization, as the angle $\theta$ between $o_A$ and $o_B$ goes to 0, on SBM networks. Y-axis shows polarization according to the corresponding method, x-axis shows $\theta$. Shaded area is 95-percentile confidence interval. Generated from 100 SBM networks.

### 4. Conclusion

APD, ADM, and TV conform to desired behavior, decreasing as the correlation between two of the three opinions increases. PC and MDS fail to do so, decreasing or staying constant as the correlation increases.

## Community consensus

This scenario models within-community opinion reinforcement and homogenisation, which are effects observed in networks where people get increasingly concentrated in echo-chambers. It's inspired by polarization as "Group Consensus", proposed by A. Bramson et al. (2017) [6]. I implement the scenario by decreasing the diversity of stances within opinions by decreasing the standard deviation of the distribution from which non-zero stances are sampled.

### 1. Desired behavior

Because stance diversity decreases, communities' stances get concentrated, and people find themselves in increasingly homogeneous echo chambers, polarization should increase.

### 2. Local scale

Like in the previous experiment, local-scale networks are fixed at size $|V| = 12$. They consist of three communities, each with an orthogonal opinion, whose stance values for their opinion are sampled from normal distributions $N(\mu, \sigma)$ with mean $\mu = 0.5$ and standard deviations $\sigma$ going from 0.2 to 0.01, as shown in Figure 14.

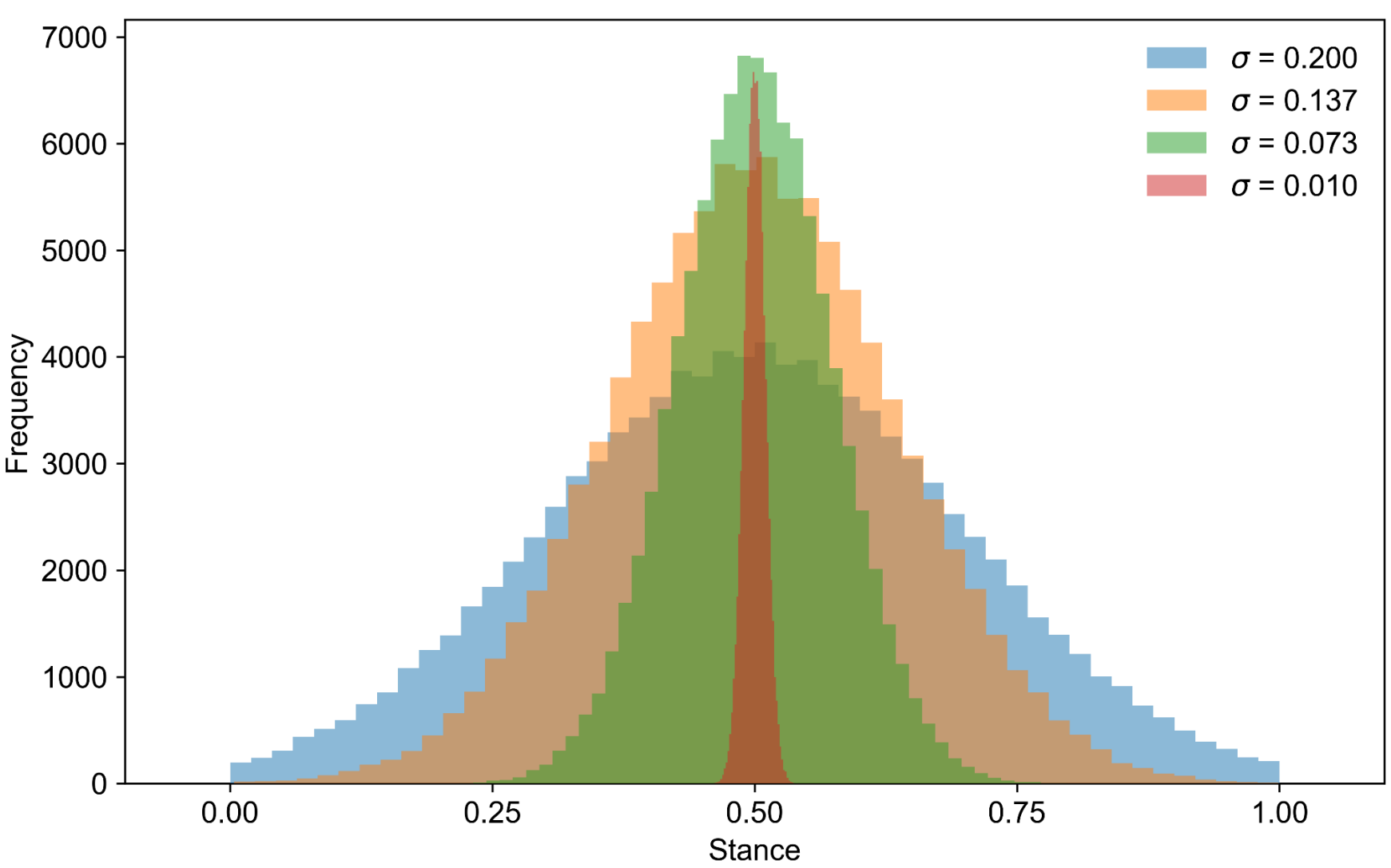

Figure 14: Example stances sampled from normal distributions with $\mu = 0.5$ and $\sigma$ going from 0.2 to 0.01. Histograms are of 100.000 stances each, limited to the interval [0, 1] in accordance with the definition of a stance.

Methods are evaluated on local-scale networks, as shown in Figure 15. Surprisingly, no method increases for any local-scale network type as $\sigma$ decreases. TV decreases in chains and in communities, as does PC. The rest stay constant or do not significantly change.

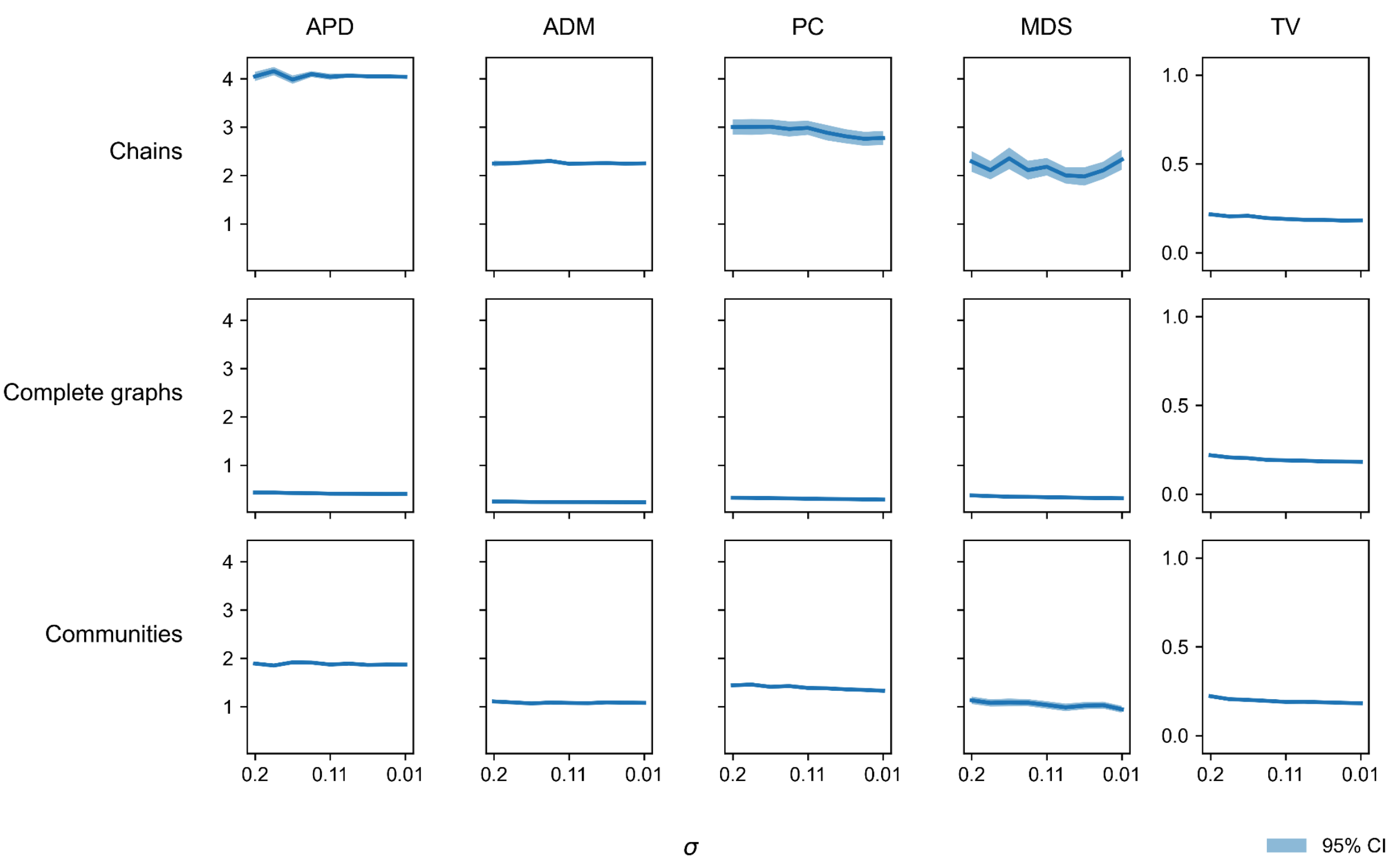

Figure 15: Polarization as $\sigma$ goes from 0.2 to 0.01 on local-scale networks. Multipolar polarization methods are on columns, and network types are on rows. Y-axis shows polarization according to the corresponding method, x-axis shows $\sigma$. The shaded areas are 95-percentile confidence intervals.

## 3. Large scale

The large-scale test functions similarly to the local-scale tests, except with SBM networks with 3 communities, each of 100 nodes. Here, as shown in Figure 16, none of the methods increase with $\sigma$ either.

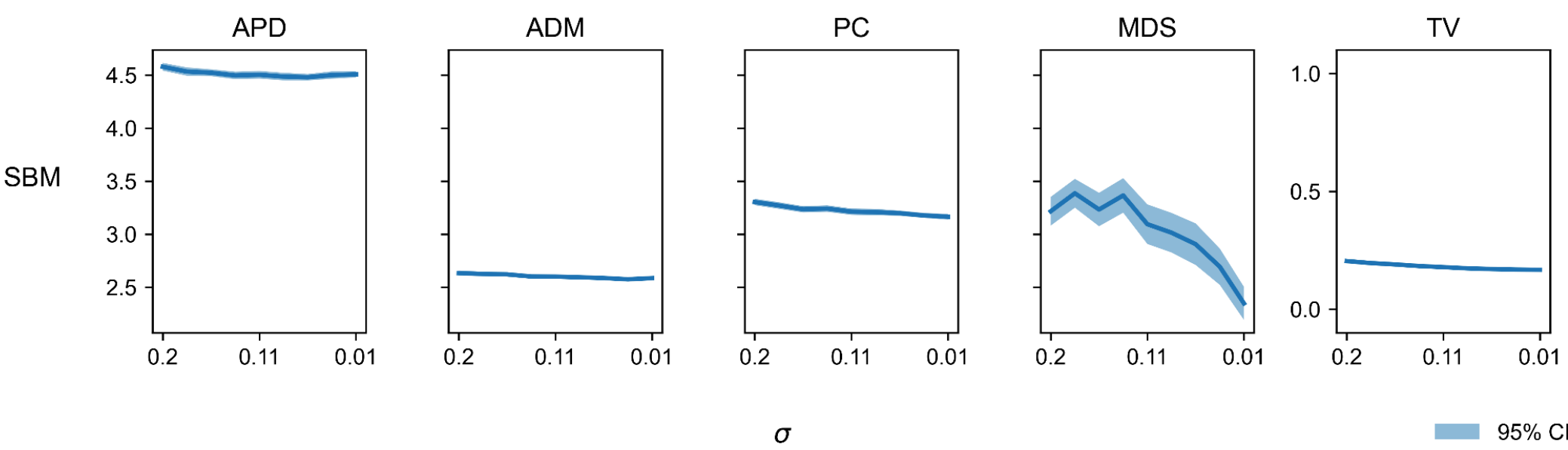

Figure 16: Polarization as $\sigma$ goes from 0.2 to 0.01 on SBM networks. Multipolar polarization methods are on columns. Y-axis shows polarization according to the corresponding method, x-axis shows $\sigma$. Shaded area is 95-percentile confidence interval from 100 generated SBM networks.

## 4. Conclusion

None of the methods conform to the desired behavior, decreasing or staying the same, as communities' opinions become less diverse.

# Discussion

Since there are so many different conceptions of what polarization is and how it can be measured, my synthetic experiments were by no means a complete list, yet they provide a useful comparison between the candidate methods on several desired behaviors. No method was perfect, as none were sensitive to the community consensus experiment, but one stands out as most suitable, as shown in Figure 17.

| Experiment | Adding nodes with unique opinions | Neutral nodes as a different community | Correlating two opinions | Community consensus |
|---|---|---|---|---|
| Desired behavior | Constant or decreasing | Constant | Decreasing | Increasing |
| APD | ✓ | ✓ | ✓ | × |
| ADM | ✓ | × | ✓ | × |
| PC | ✓ | ✓ | × | × |
| MDS | × | × | × | × |
| TV | × | × | ✓ | × |

Figure 17: Summary of whether each method conforms to desired behavior in the respective synthetic experiments.

Based on my experimentation and interpretation of metric behaviors, Average Pairwise Distance, proposed by Hohmann, et al. (2023) [12, pp. 22–23], is the best candidate method for quantifying multipolar polarization since:

- When unique orthogonal opinions are added, it decreases as expected.
- When a neutral community is arbitrarily interpreted as having a unique orthogonal opinion, it stays constant as expected.
- When two of three network opinions become correlated, it decreases as expected.

There are several limitations to these comparisons, including the generated networks; SBMs don’t emulate real-world social networks perfectly, partly since they’re not transitive – the friend of my friend is (not) more likely to be my friend. A key feature of real networks is the power law degree distribution, which the SBMs don’t have.

Some limitations of Average Pairwise Distance as a distance metric are inherited from generalized Euclidean distance; it isn’t normalized, and it doesn't incorporate many aspects of polarization, including affective polarization. More work is needed to show why it isn’t sensitive to community consensus. This could also be a feature of generalized Euclidean distance.

A. Bramson et al. (2017) [6] outline 9+ senses of polarization, some of which could be the basis for further experimentation, including Size Parity – communities being of equal size and Community Fracturing – larger communities splitting into distinct subcommunities. Furthermore, my experiments were conducted independently rather than simultaneously, without considering combinations of changes to the network, which could provide additional useful insights.

# Code & Contributions

Code to reproduce all findings and synthetic data visualizations is accessible in the GitHub repository [20] or by request.

Thanks to my supervisor, Michele Coscia, for providing code at the start of the project, including resources for calculating generalized Euclidean distance, APD, TV, and more. Where the code was not written/modified by me, it is indicated in the files.

# References

[1] S. Iyengar, Y. Lelkes, M. Levendusky, N. Malhotra, and S. J. Westwood, “The Origins and Consequences of Affective Polarization in the United States,” *Annual Review of Political Science*, vol. 22, no. Volume 22, 2019, pp. 129–146, May 2019, doi: 10.1146/annurev-polisci-051117-073034.

[2] M. P. Fiorina and S. J. Abrams, “Political Polarization in the American Public,” *Annual Review of Political Science*, vol. 11, no. Volume 11, 2008, pp. 563–588, Jun. 2008, doi: 10.1146/annurev.polisci.11.053106.153836.

[3] J.-M. Esteban and D. Ray, “On the Measurement of Polarization,” *Econometrica*, vol. 62, no. 4, pp. 819–851, 1994, doi: 10.2307/2951734.

[4] P. Bauer, *Conceptualizing and measuring polarization: A review*. 2019. doi: 10.31235/osf.io/e5vp8.

[5] M. Gestefeld, J. Lorenz, N. T. Henschel, and K. Boehnke, “Decomposing attitude distributions to characterize attitude polarization in Europe,” *SN Soc Sci*, vol. 2, no. 7, p. 110, Jul. 2022, doi: 10.1007/s43545-022-00342-7.

[6] A. Bramson *et al.*, “Understanding polarization: Meanings, measures, and model evaluation,” *Philosophy of science*, vol. 84, no. 1, pp. 115–159, 2017.

[7] F. Baumann, P. Lorenz-Spreen, I. M. Sokolov, and M. Starnini, “Modeling Echo Chambers and Polarization Dynamics in Social Networks,” *Phys. Rev. Lett.*, vol. 124, no. 4, p. 048301, Jan. 2020, doi: 10.1103/PhysRevLett.124.048301.

[8] F. Baumann, P. Lorenz-Spreen, I. M. Sokolov, and M. Starnini, “Emergence of Polarized Ideological Opinions in Multidimensional Topic Spaces,” *Phys. Rev. X*, vol. 11, no. 1, p. 011012, Jan. 2021, doi: 10.1103/PhysRevX.11.011012.

[9] R. Interian, R. G. Marzo, I. Mendoza, and C. C. Ribeiro, “Network polarization, filter bubbles, and echo chambers: an annotated review of measures and reduction methods,” *Int Trans Operational Res*, vol. 30, no. 6, pp. 3122–3158, Nov. 2023, doi: 10.1111/itor.13224.

[10] M. Coscia, “Generalized Euclidean measure to estimate network distances,” in *Proceedings of the international AAAI conference on web and social media*, 2020, pp. 119–129. Accessed: Mar. 21, 2024. [Online]. Available: https://aaai.org/ojs/index.php/ICWSM/article/view/7284

[11] M. Coscia, A. Gomez-Lievano, J. Mcnerney, and F. Neffke, “The Node Vector Distance Problem in Complex Networks,” *ACM Comput. Surv.*, vol. 53, no. 6, pp. 1–27, Nov. 2021, doi: 10.1145/3416509.

[12] M. Hohmann, K. Devriendt, and M. Coscia, “Quantifying ideological polarization on a network using generalized Euclidean distance,” *Sci. Adv.*, vol. 9, no. 9, p. eabq2044, Mar. 2023, doi: 10.1126/sciadv.abq2044.

[13] I. Borg and P. J. F. Groenen, *Modern Multidimensional Scaling*. in Springer Series in Statistics. New York, NY: Springer, 2005. doi: 10.1007/0-387-28981-X.

[14] “2.2. Manifold learning,” scikit-learn. Accessed: May 12, 2024. [Online]. Available: https://scikit-learn/stable/modules/manifold.html

[15] D. J. Klein and M. Randić, “Resistance distance,” *J Math Chem*, vol. 12, no. 1, pp. 81–95, Dec. 1993, doi: 10.1007/BF01164627.

[16] S. Martin-Gutierrez, J. C. Losada, and R. M. Benito, “Multipolar social systems: Measuring polarization beyond dichotomous contexts,” *Chaos, Solitons & Fractals*, vol. 169, p. 113244, 2023.

[17] P. W. Holland, K. B. Laskey, and S. Leinhardt, “Stochastic blockmodels: First steps,” *Social Networks*, vol. 5, no. 2, pp. 109–137, Jun. 1983, doi: 10.1016/0378-8733(83)90021-7.

[18] E. Abbe and C. Sandon, “Community detection in general stochastic block models: fundamental limits and efficient recovery algorithms.” arXiv, Apr. 04, 2015. doi: 10.48550/arXiv.1503.00609.

[19] A. A. Amini and E. Levina, “On semidefinite relaxations for the block model.” arXiv, Mar. 16, 2016. doi: 10.48550/arXiv.1406.5647.

[20] C. Weidemann, “Quantifying Multipolar Polarization.” Feb. 2024. Accessed: May 08, 2024. [Online]. Available: https://github.com/Christian-Weidemann/Quantifying-Multipolar-Polarization